\documentclass[%
 reprint,
superscriptaddress,
twocolumn,
amsmath,
amssymb,
aps,
prb,
longbibliography,
]{revtex4-2}
\usepackage{graphicx}
\usepackage{dcolumn}
\usepackage{bm}
\usepackage[colorlinks=true,linkcolor=blue,citecolor=blue,urlcolor=blue,breaklinks]{hyperref}
\usepackage{amsmath}  
\usepackage{amssymb}
\usepackage{xcolor}
\usepackage{braket}

\begin{document}

\title{Convergence of sample-based quantum diagonalization on a variable-length cuprate chain}

\author{L. Andrew Wray}
\author{Cheng-Ju Lin}
\author{Vincent Su}
\author{Hrant Gharibyan}
\affiliation{BlueQubit Inc., San Francisco, CA 94105, USA}

\begin{abstract}

Sample-based quantum diagonalization (SQD) is an algorithm for hybrid quantum-classical molecular simulation that has been of broad interest for application with noisy intermediate scale quantum (NISQ) devices. However, SQD does not always converge on a practical timescale. Here, we explore scaling of the algorithm for a variable-length molecule made up of 2 to 6 copper oxide plaquettes with a minimal molecular orbital basis. The results demonstrate that enabling all-to-all connectivity, instituting a higher expansion order for the SQD algorithm, and adopting a non-Hartree-Fock molecular orbital basis can all play significant roles in overcoming sampling bottlenecks, though with tradeoffs that need to be weighed against the capabilities of quantum and classical hardware. Additionally, we find that noise on a real quantum computer, the Quantinuum H2 trapped ion device, can improve energy convergence beyond expectations based on noise-free statevector simulations.

\end{abstract}

\date{\today}
\maketitle

\twocolumngrid

\begin{figure}
    \centering \includegraphics[width=0.5\textwidth]{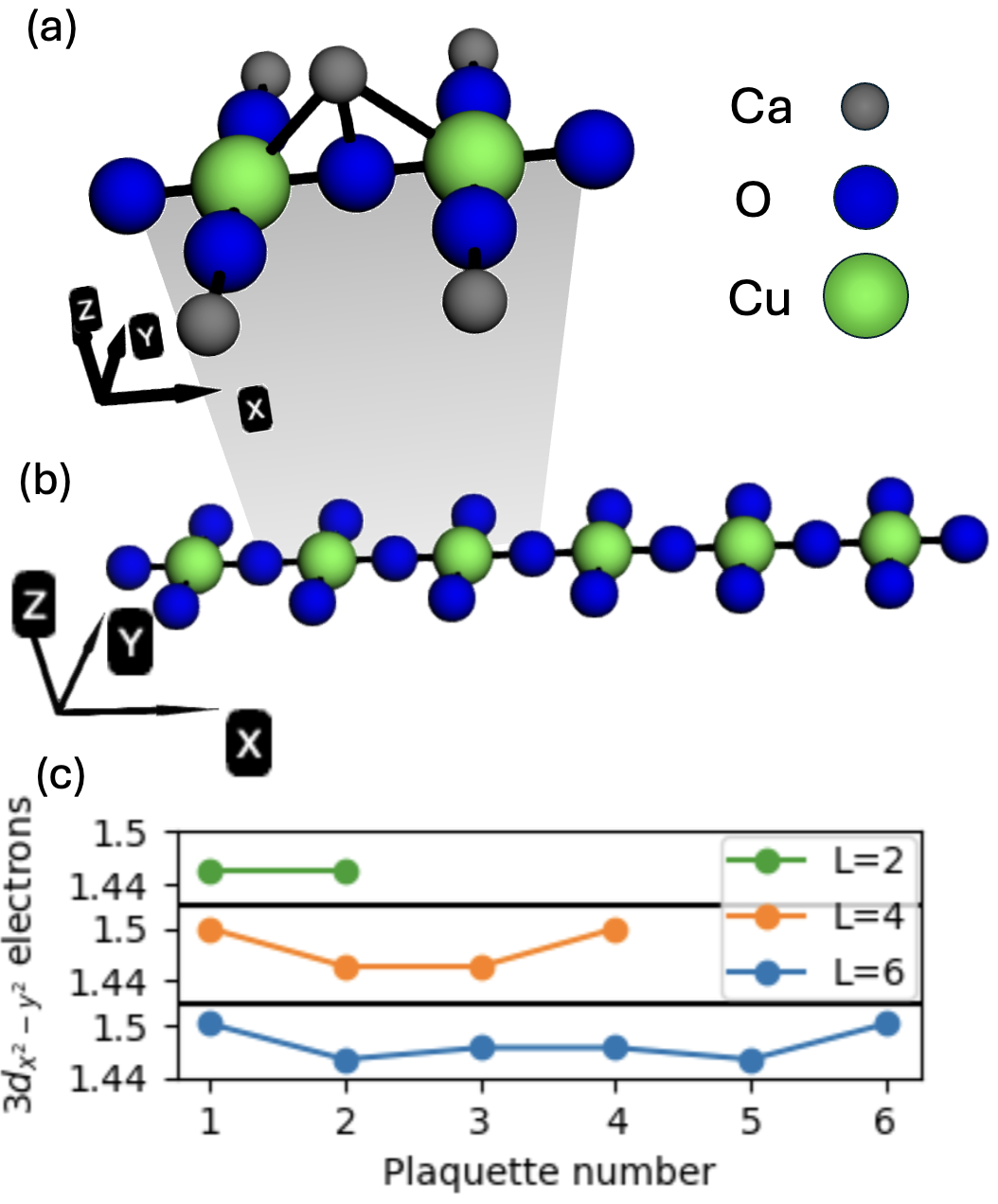}
    \caption{\textbf{Cuprate chain model system}: (a) Diagram of the charge-neutral chain molecule used for Hatree-Fock calculations. (b) The simplified and extended cuprate chain with L=6 copper oxide plaquettes. (c) Charge density distribution in Cu $\text{3d}_{s^2-y^2}$ orbitals as a function of chain length.}
    \label{fig:fig1}
\end{figure}

\begin{figure*}
    \centering \includegraphics[width=0.95\textwidth]{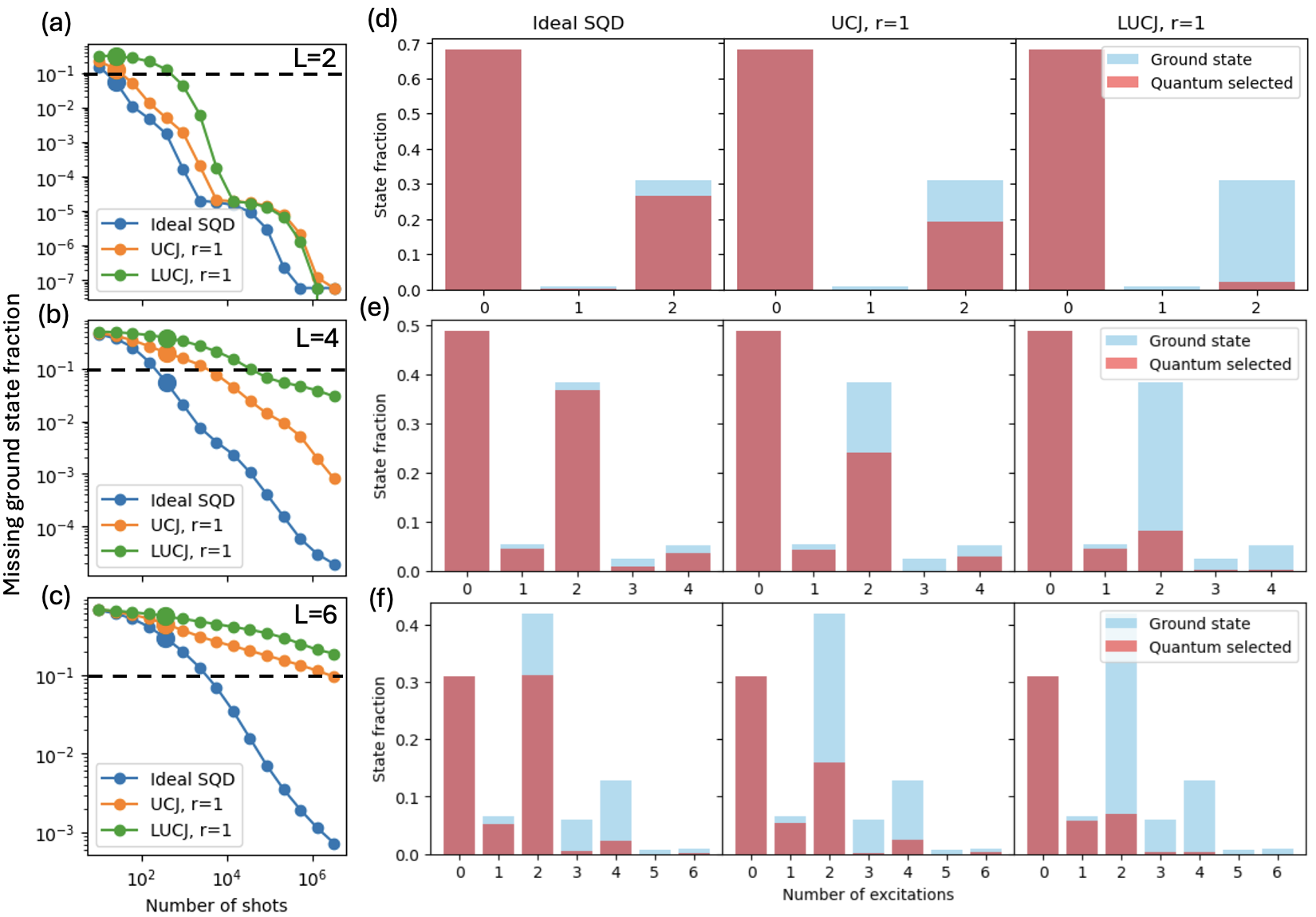}
    \caption{\textbf{SQD versus system size}: (a-c) Fraction of the ground state projecting onto SQD determinants as a function of shot number for (a) L=2, (b) L=4, and (c) L=6 plaquette length chains. (d-f) Blue bars show the full ground state decomposed by excitation number on the Hartree-Fock basis. The fraction accounted for within the three quantum-selected bases is shown in red, for shot numbers indicated with large dots in panels (a-c). All plots show expectation values rather than single-experiment outcomes, as described in Section III of the SM.}
    \label{fig:fig2}
\end{figure*}

Sample-based quantum diagonalization (SQD) algorithms are a class of hybrid quantum-classical computational techniques proposed in 2023 that make use of quantum computing devices to approximate a desired many-electron state \cite{IBM_LUCJ0,IBM_LUCJ1,IBM_LUCJ2,QSCI,UCJ_HeadGordon2025,IBM_LUCJ_periodic2025,fatal_flaw}. This wavefunction is measured to identify Slater determinants for a selected configuration interaction (SCI) calculation on classical computing hardware. Wavefunction emulation on a quantum device conveys the advantage that Slater determinants are not limited to specific excitation number sectors as in fully classical approaches such as coupled cluster and heat-bath configuration interaction (HCI) \cite{HCI0} modeling. Recent work has demonstrated the application of SQD on molecules as large as the [4Fe-4S] iron-sulfur cluster model system, with 54 electrons in an active space of 36 orbitals, using IBM quantum devices \cite{IBM_LUCJ0,IBM_LUCJ1}. However, the computational complexity of SQD is not yet well characterized for computationally expedient algorithms such as the unitary cluster Jastrow ansatz (UCJ) and local UCJ ansatz (LUCJ) \cite{LUCJ1,IBM_LUCJ0,IBM_LUCJ1}, which can result in less compact determinant sets and convergence plateaus \cite{fatal_flaw}. Here, we investigate techniques to overcome convergence challenges and demonstrate that the observed computational complexity of SQD sampling on chain molecules of different lengths appears to be compatible with computationally efficient application to large chemical systems.

A corner sharing copper oxide chain is selected as a model many-body system featuring strong entanglement. Representing this entanglement is expected to call for a rapid growth in the excitation number allowed within classical modeling approach such as coupled cluster (CC) calculations as the chain length is increased. Though the CC excitation order needed for accurate modeling of \emph{any} molecular system is expected to grow with the molecule size, our goal is to make this phenomenon visible using a minimal orbital basis so that it can be traced to assess consequences for the computational complexity of SQD calculations.

A first principles Hartree-Fock calculation for a charge neutral dimer molecule with formula Ca$_5$Cu$_2$O$_7$ (Fig.~\ref{fig:fig1}(a)) is used to populate 1- and 2-body Hamiltonian terms for the widely referenced cuprate ``two-band" model (c.f. \cite{twoBand1, twoBand2, twoBand3, ZhangRice1988, twoBand_Sawatzky}). This minimal modeling basis consists of two molecular orbitals per CuO$_3$ cell representing the strongly correlated copper 3d$_{x^2-y^2}$ atomic orbital, as well as a superposition of $\sigma$-bonding oxygen 2p atomic orbitals. Details of the model construction and first principles calculation are found in SI Section A. 

The two-band Hamiltonian is then used to construct a plaquette chain featuring on-site and nearest neighbor interplaquette interactions, as shown in Fig.~\ref{fig:fig1}(b). The resulting ground state Cu 3d$_{x^2-y^2}$ charge density distribution from a full configuration interaction (FCI) calculation is roughly uniform and matches expectations for Mott insulating cuprate chain compounds in the SrCuO$_2$ family (3d$_{x^2-y^2}$ charge density $n_{x^2-y^2} \sim 1.4$ \cite{XPS_nd_Kotani1997}), which is within the $1.2 \lesssim n_{x^2-y^2} \lesssim 1.5$ range expected for Mott insulating cuprates \cite{Cuprate_nd_NMR1, Cuprate_nd_NMR2}. Some additional charge density is expected due to the finite size of our chain systems, which feature a terminal oxygen atom that breaks the CuO$_3$ stoichiometry of an infinite chain.

Nearest neighbor spin orientations are negatively correlated, as expected for a Mott insulating cuprate. The kinetic Hamiltonian of the model is tuned to roughly align spin superexchange with SrCuO$_2$ (see details in Section II in the Supplementary Material, SM), which is an antiferromagnetic Heisenberg spin chain compound with exchange constant J$=$0.23 eV identified from direct measurement of the spinon dispersion \cite{CKim_spinons_2006}. Significantly, the lowest energy spin excitation for each chain falls between 0.1 to 0.2 eV, which is greater than the standard chemical accuracy threshold of E$_C$=27 meV (1 mHa), emphasizing the importance for a numerical model to accurately represent entanglement within the 2$^L$ spin configuration basis of the chain.

The efficiency of SQD in describing the ground states of these chains is explored in Fig. 2. Fig. 2(a-c) show the fraction '$f$' of the exact many-body ground state ($|g\rangle$) that is missing after projecting onto the SQD basis of sampled determinants ($|i\rangle$), as a function of the number of measurements (shots) performed on a quantum device:
\begin{equation}
f = 1 -  \Sigma_i \langle g|i\rangle \langle i |g\rangle
\end{equation}

Each shot represents the measurement of a single Slater determinant; however, many measurements yield previously identified determinants that do not enlarge the basis set for diagonalization. Curves are plotted for (orange) UCJ, a practical SQD ground state approximation algorithm involving all-to-all connectivity, and (green) LUCJ, a version of UCJ in which interaction terms are truncated to the local honeycomb connectivity on current IBM quantum devices. Both approaches are performed with only the lowest order (r=1) expansion of the UCJ operator, consistent with earlier studies that have primarily used r=1 to 2 \cite{QSCI,UCJ_HeadGordon2025,fatal_flaw,IBM_LUCJ0,IBM_LUCJ1,IBM_LUCJ2}. An additional benchmarking curve is provided for (blue) `ideal SQD', which is not achieved with a quantum circuit but rather by sampling from a perfect representation of the ground state obtained via full configuration interaction (FCI) calculations. 

\begin{figure*}
    \centering \includegraphics[width=1.0\textwidth]{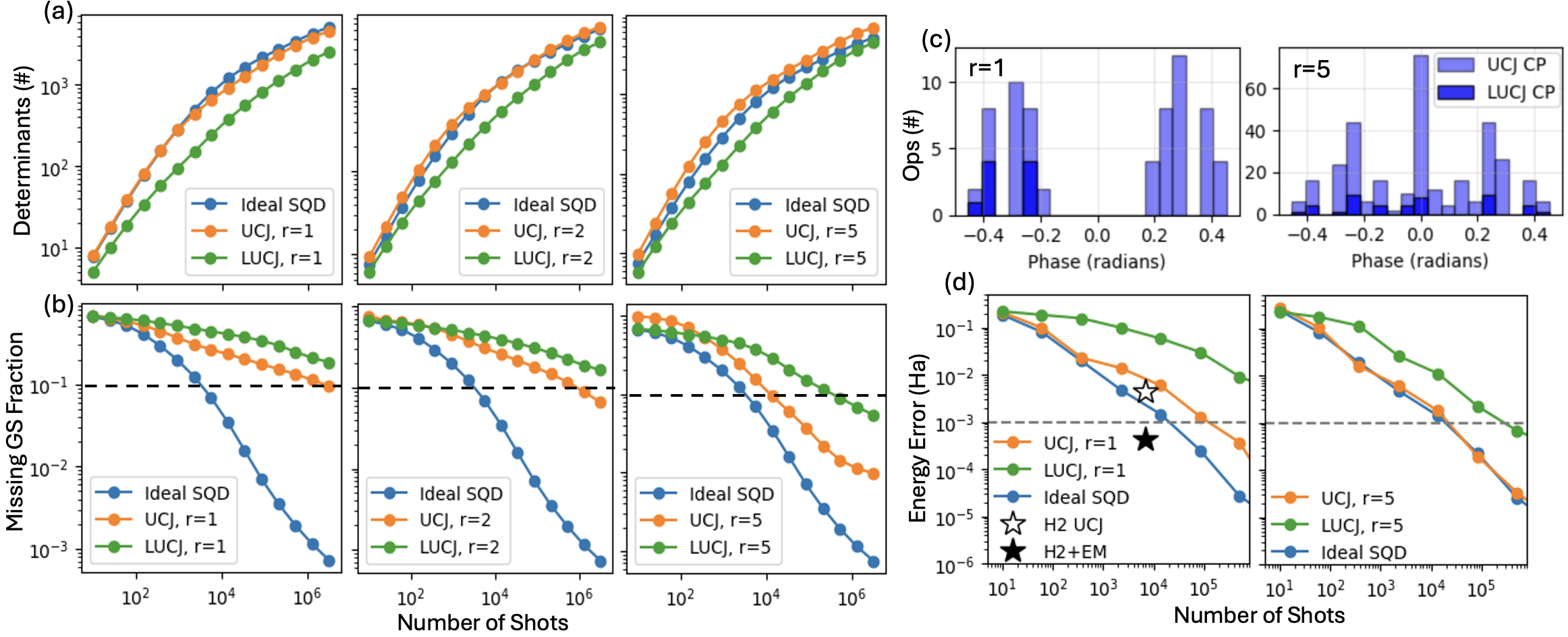}
    \caption{\textbf{UCJ operator expansion order}: The performance of SQD with different orders of double-factorized operator expansion is shown for an L=6 site chain. The (a) number of determinants and (b) fraction of the ground state lost when projecting to these determinants are plotted against shot number. (c) A histogram of cp gate amplitudes, which represent Coulomb interaction terms. (d) Ground state energy predictions from SQD. Data from a 7000 shot run of the r=1 UCJ cirquit on Quantinuum H2 hardware are included for (higher energy) unmodified and (lower energy) error mitigated versions of the output. Emulated circuit results in panels (a,b,d) represent mult-run averages and expectation values rather than single-experiment outcomes, as described in Section III of the SM.}
    \label{fig:fig3}
\end{figure*}

\begin{figure}[h]
    \centering \includegraphics[width=0.5\textwidth]{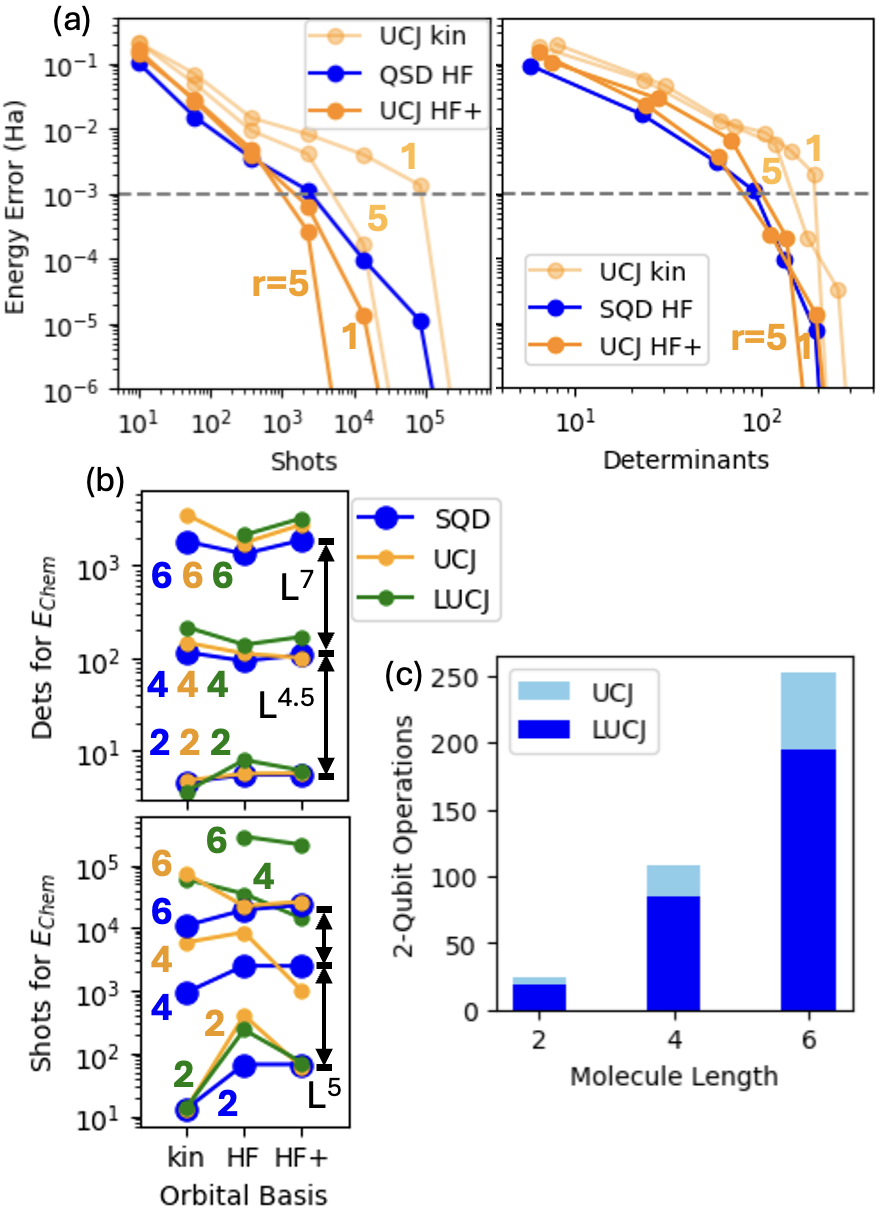}
    \caption{\textbf{Computational resources for chemical accuracy}: (a, left) Energy error versus shot number for emulated UCJ on the L=4 chain with (light orange) kinetic orbitals, and (dark orange) HF+ orbitals is compared with (blue) ideal SQD with HF orbitals. (a, right) Energy error as a function of the number of unique Slater determinants. (b, top) Number of unique determinants in each orbital basis at chemical accuracy for length L=2, 4, and 6 chains with r=5 operator expansions. Ideal SQD is plotted with a larger point size for visibility. Black brackets indicate the vertical spacings expected for $L^{4.5}$ and $L^7$ growth rates, which roughly match gaps between the ideal SQD data points. (b, bottom) Shots required to reach chemical accuracy. Black brackets indicate vertical spacings expected for $L^5$ complexity, and are aligned with HF+ basis data points for ideal SQD. (c) The number of two qubit operations (XX+YY and CP gates) involved in each calculation.}
    \label{fig:fig4}
\end{figure}

As might be expected, the ground state is always most accurately represented by ideal SQD, followed successively by UCJ and lastly the truncated LUCJ. These three curves show qualitatively similar trends for the shortest chain (L=2, Fig.~\ref{fig:fig2}(a)), but diverge for longer chains, with a plateauing trend observed for UCJ and LUCJ. A similar convergence plateau was noted for r=2 UCJ and LUCJ modeling of N$_2$ in Ref.~\cite{fatal_flaw} and proposed to constitute a potential bottleneck for practical SQD, though follow-up investigations have achieved superior convergence \cite{IBM_LUCJ_Motta_r2024}. For the L=6 chain, UCJ takes roughly 3 orders of magnitude longer than ideal SQD to reach 90$\%$ coverage of the ground state, and LUCJ trails by 1-2 further orders of magnitude (Fig.~\ref{fig:fig2}(c)).

To better understand the SQD output, we decompose the projected ground state by excitation number, yielding the bar plots in Fig.~\ref{fig:fig2}(d-f). Here, excitation number refers to the number of occupied Hartree-Fock molecular orbitals above the Fermi level for a given Slater determinant. For good visual contrast, sampling is terminated at low shot numbers indicated by enlarged data points in Fig.~\ref{fig:fig2}(a-c). In all cases, the ground state fraction with low excitation number ($n_{ex}<3$) is well accounted for due to having high wavefunction weight and low total multiplicity. Excitation pairs are more common than single electron excitations within the ground state, reflected in the high weight within all even excitation-number bins relative to their immediate neighbors. Determinants with larger excitation number ($n_{ex}>2$) are nonetheless significant and account for $\gtrsim 10 \%$ of the ground state for the L=2 and 4 chains. It should be noted that the contribution of these states to the ground state energy can in some cases be large relative to their fractional representation within these bar graphs \cite{pertTheoryDeriv}. The UCJ and LUCJ bases show reduced coverage of high-excitation-number states, but are not readily distinguished from ideal SQD in the $n_{ex}<$2 excitation sector.

A straightforward approach to overcome the convergence plateaus observed in Fig.~\ref{fig:fig2}(b-c) is to simply increase the expansion order (r) of the UCJ operator. As the expansion order is increased, the UCJ operator provides a successively more accurate representation of electron-electron interaction terms in the Hamiltonian. Fig.~\ref{fig:fig3}(a) examines the largest L=6 chain case and shows that increasing the expansion order causes UCJ and LUCJ to yield a larger number of determinants. However, the improvement in the number of unique determinants seen when increasing from the lowest expansion order r=1 to r=5 is comparable to simply increasing shot number by less than an order of magnitude. This is a dubious advantage when taken on its own, as the number of quantum gates required for the algorithm scales linearly with r.

A more pronounced improvement can be seen in the ground state fraction versus shot number curves in Fig.~\ref{fig:fig3}(b). Increasing the UCJ order from r=1 to r=5 largely eliminates the plateau and causes the UCJ trend curve to strongly resemble ideal SQD. The rate of convergence to 90$\%$ ground state projection is improved by more than two orders of magnitude. In terms of the rate of energy convergence, increasing r from 1 to 5 improves UCJ by a factor of $\sim$8 on the shot number axis near the chemical accuracy threshold and renders it almost indistinguishable from ideal SQD. Shot number for LUCJ improves more than an order of magnitude, easily justifying the factor of 5 growth in the number of quantum gates on a sufficiently stable quantum device. One reason may be that the LUCJ truncation removes the great majority of the controlled phase (CP) operators through which electron-electron interactions are enacted (see Fig.~\ref{fig:fig3}(c)), making the r=1 LUCJ operator a very sparse operation with just 11 CP terms. 

A motivational concept behind the LUCJ algorithm is that many of the CP terms truncated away may represent longer-range interactions that are relatively small and unimportant. The histogram of r=5 CP gate phase amplitudes in Fig.~\ref{fig:fig3}(c,right) shows that there is indeed a peak at 0, which contains a large number of terms with phase amplitudes smaller than 1$\%$ of the dynamic range in the plot. This peak would likely be significantly larger if our interacting Hamiltonian were not limited to same- and nearest-neighbor- plaquettes due to originating from first principles calculations for a two-plaquette system. These low-amplitude terms account for 17$\%$ of the LUCJ CP operators and 23$\%$ of the UCJ operators, consistent with LUCJ preferentially retaining larger-amplitude terms. However, it is also easy to see that much of the truncation within LUCJ occurs for larger amplitude terms.

To demonstrate the feasibility of these circuits on near-term quantum hardware, the energy error from a 7000 shot r=1 UCJ circuit run with all-to-all connectivity on the Quantinuum H2-2 quantum computer~\cite{QuantinuumH2_2} is plotted with two black stars in Fig.~\ref{fig:fig3}(d, left). The higher energy value (hollow star) represents a single 7000-shot measurement, for which 35$\%$ of output determinants have the correct electron number in both spin sectors. This extrapolates to $\sim$28$\%$ of the determinants being truly error free. Additional details of the H2-2 experiment can be found in Section I of the SM. The solid black star at lower energy is obtained through coarse error mitigation applied to bitstrings with the wrong electron number in a given spin sector. This is achieved by creating electrons or holes (flipping bits) in single particle states for which the change will help the most to align with mean values within the full dataset, drawing inspiration from earlier work \cite{IBM_LUCJ1}. The Quantinuum H2-2 calculation energy with unmitigated quantum device output is consistent with expectations based on the error distribution for single measurements, however it is surprising to see that the error mitigated ground state energy is significantly better than noise-free UCJ.

A partial explanation for the significantly improved ground state energy estimate following error mitigation may lie in the number of determinants generated, as well as the small number of orbital degrees of freedom in the simulation. The error mitigated basis contains 1876 unique determinants, roughly corresponding to the number expected from a far longer 80,000 shot experiment (compare with Fig.~\ref{fig:fig2}(a)). The error mitigated calculation energy actually exceeds expectations for an 80,000 shot UCJ circuit and approaches the level expected for ideal UCJ at this shot number. A similar improvement upon error mitigation is seen for the L=6 r=2 circuit (see Fig S2 of the SM). This phenomenon indicates that in forcing noise-impacted measurements into relatively high-probability states with the correct electron number, the error mitigation algorithm is generating a number of complementary determinants that are difficult to access directly from UCJ but have high relevance to the ground state energy. We note that an improvement of error mitigated LUCJ versus error-free LUCJ can also be seen in the simulations within Ref.~\cite{LUCJ_calc_vs_EM2025}, though the difference falls within the standard deviation for repeated calculations.

As a final approach to improve convergence, the performance of UCJ on different molecular orbital bases is explored in Fig.~\ref{fig:fig4}. Previous UCJ calculations have made use of the Hartree-Fock basis, but have noted that sampling efficiency can be improved by performing an orbital basis transformation on the circuit output before measurement \cite{IBM_LUCJ1,IBM_LUCJ2} or by artificially applying a square root operation to wavefunction amplitudes in an emulated UCJ circuit \cite{fatal_flaw}. As a more controlled approach that retains the full structure of the UCJ algorithm, we have instead performed the standard calculation on alternative molecular orbital bases that are expected to be more strongly perturbed by the interacting Hamiltonian, resulting in the more rapid generation of determinants. These bases are (kin) the eigenstates of the kinetic Hamiltonian, and (HF+) a basis that is \emph{overcorrected} for the mean-field Coulomb interactions considered in Hartree-Fock. (see details in Section IV of the SM)

As expected, both of these alternate molecular orbital bases result in more unique determinants as a function of shot number, and can in certain cases improve the speed of convergence to chemical accuracy (see L=2 case in Fig.~\ref{fig:fig4}(a, left)). However, this comes at the cost of a less compact basis set. As seen in Fig.~\ref{fig:fig4}(a, right), the number of determinants required for a given energy error is generally higher, meaning that the classical diagonalization cost is likely to be higher unless special measures are taken to compress the basis \cite{handover_VQE}, or diagonalization is performed on a different basis from the UCJ calculation as in Ref.~\cite{IBM_LUCJ1, IBM_LUCJ2}. The number of determinants required for chemical accuracy in all r=5 circuits is shown in Fig.~\ref{fig:fig4}(b, top), and shows mostly v-shaped contours for L=4 and 6 calculations, consistent with the HF molecular orbital basis providing better compactness for those larger systems. 

By contrast, the number of shots needed to achieve chemical accuracy shows a greater diversity of trends (Fig.~\ref{fig:fig4}(b,bottom). Ideal SQD converges fastest in the kin basis, while UCJ tends to do well in the HF+ basis. Bars at right show the gaps expected between L=$2\rightarrow 4$ and L=$4\rightarrow 6$ calculations with L$^{5}$ scaling, which roughly matches the trend for ideal SQD. Combining this with the $N_{op}\sim L^{\sim2.2}$ operator count scaling seen in Fig.~\ref{fig:fig4}(c) gives an observed computational complexity of roughly L$^7$, or n$^7$ for an n-qubit calculation. We note that the necessary expansion order $r$ is expected to saturate for large systems, as it represents the number of orbital bases required to provide diagonalized representations of Coulomb interactions, which are intrinsically local.

If this trend holds for other and larger systems, it is similar or superior to the n$^{8}$ diagonalization time of configuration interaction calculations such as CISDT that allow up to 3 excitations \cite{comp_complexity}, suggesting that SQD approaches may be computationally efficient for high excitation-order determinant selection. However, the number of unique determinants required for chemical accuracy does not adhere to a fixed polynomial scaling order within our data (see Fig.~\ref{fig:fig4}(b,top)), consistent with the idea that many-body ground states are inefficient to represent on classical computers. A greater-than-polynomial asymptotic trend in the number of required determinants would imply that the favorable L$^7$ scaling of QSD sampling time will eventually break down, just as larger systems call for higher excitation order in conventional CC and CI calculations.

In conclusion, we have emulated UCJ and LUCJ circuits for cuprate chains of variable length, as well as simulating an ideal SQD algorithm that samples from the exact many-body ground state. Using the low order (r$\le$2) UCJ and LUCJ operators favored in previous studies shows a plateauing trend in sampling on larger systems, similar to that seen with r=2 UCJ and LUCJ for the N$_2$ molecule in Ref.~\cite{fatal_flaw}. However, successful convergence of LUCJ N$_2$ energies to chemical accuracy was later achieved by increasing r to values of 4-8 \cite{IBM_LUCJ_Motta_r2024}, and we find that our convergence can likewise be significantly improved in several ways. Critically, increasing the expansion order from r=2 to 5 alone is sufficient to overcome plateauing behavior in our simulations and cause UCJ convergence to closely resemble ideal SQD, though LUCJ continues to require greater computational time. Further, we show that alternate non-Hartree-Fock molecular orbital bases can speed up the generation of determinants. In the context of natural tradeoffs between connectivity, fidelity, and clock speed within current quantum computers, these factors favor systems with all-to-all connectivity and better fidelity to enable a more accurate and higher order representation of the interacting Hamiltonian. 

Experimental validation of the six-plaquette (L=6) chain UCJ operator on a real quantum device (the Quantinuum H2-2 quantum computer) yields a surprisingly improved sampling time relative to the same circuit emulated with perfect gate fidelity, suggesting that error-mitigated noise can provide a complementary form of quantum sampling. Collectively, these results provide an array of approaches for improving convergence in UCJ and LUCJ, and demonstrate that these SQD sampling approaches can achieve competitive computational scaling for a highly entangled cuprate spin chain system, suggesting promise for application to larger molecules and systems with strong electronic correlations.

\bibliography{ref}
\clearpage

\setcounter{equation}{0}
\setcounter{figure}{0}
\setcounter{table}{0}
\setcounter{page}{1}
\makeatletter
\renewcommand{\thesection}{S\arabic{section}}
\renewcommand{\theequation}{S\arabic{equation}}
\renewcommand{\thefigure}{S\arabic{figure}}

\onecolumngrid
\begin{center}
\textbf{\large Supplemental Material for: Convergence of sample-based quantum diagonalization on a variable-length cuprate chain}
\\~\\
L. Andrew Wray$^{1,2}$, Cheng-Ju Lin$^{3,4}$, Vincent Su$^{1}$, Hrant Gharibyan$^{1,2}$ \\
\vspace{.05in}
\small{
$^{1}$\textit{BlueQubit Inc., San Francisco, CA 94105, USA}
}\\

\end{center}

\title{Supplemental Material for: Convergence of sample-based quantum diagonalization on a variable-length cuprate chain}

\date{\today}

\author{L. Andrew Wray}
\author{Cheng-Ju Lin}
\author{Vincent Su}
\author{Hrant Gharibyan}
\affiliation{BlueQubit Inc., San Francisco, CA 94105, USA}

\maketitle

\onecolumngrid

\section{UCJ Circuit:}

\begin{figure}[ht]
    \centering \includegraphics[width=1.0\textwidth]{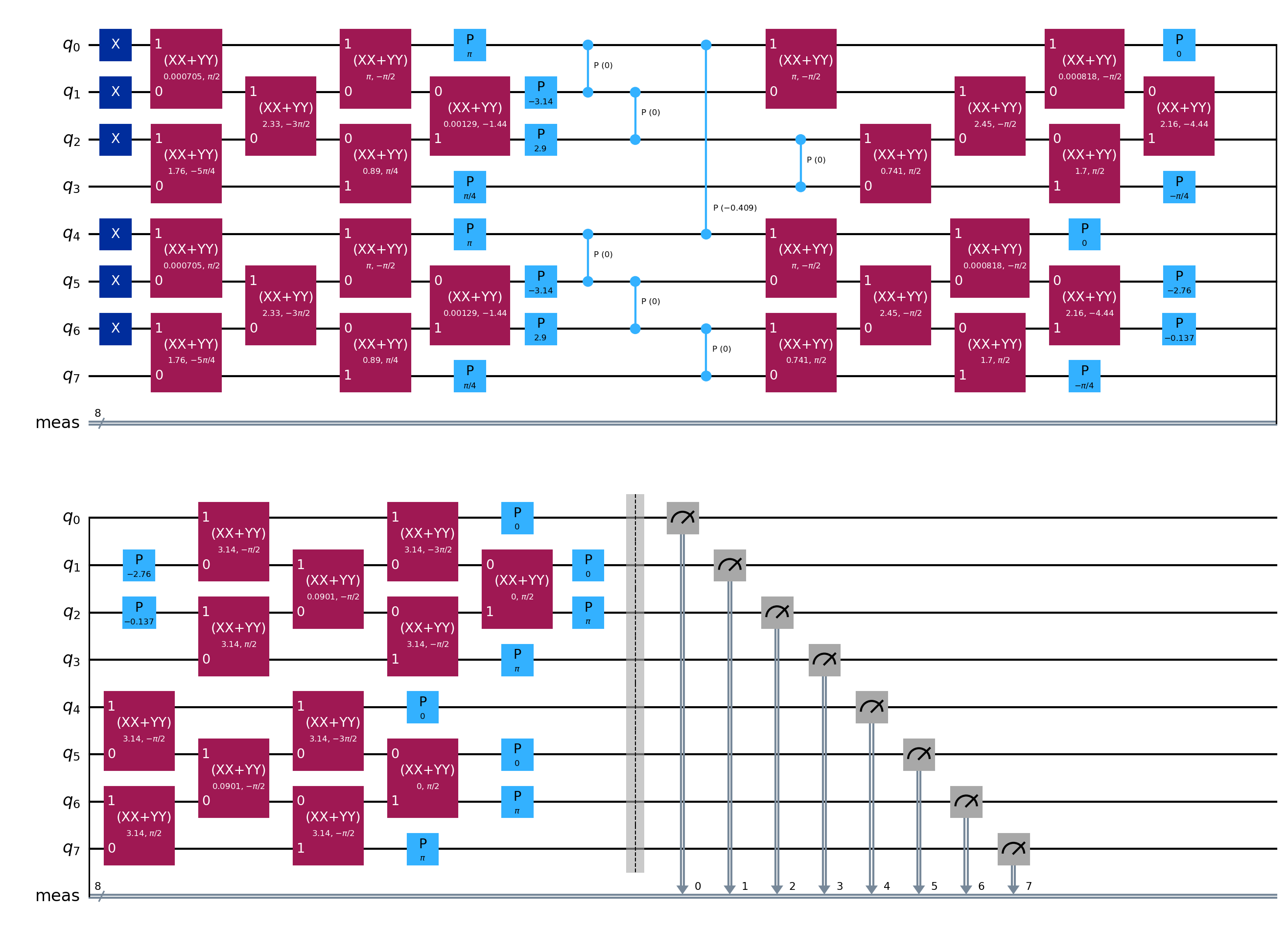}
    \caption{\textbf{Quantum circuit for LUCJ.} The quantum circuit for LUCJ on the L=2 chain with 8 spin-orbitals. A detailed tutorial for simulating UCJ and LUCJ circuits in qiskit can be found in Ref.~\cite{UCJ_tutorial}.}
    \label{fig:figS1}
\end{figure}

A quantum circuit for r=1 LUCJ on the L=2 chain is shown in Fig.~\ref{fig:figS1}. The system contains 4 orbitals which are mapped to 8 qubits, with spin=up orbital occupancy indicated by the state of qubits 0-3 and spin down in 4-7. 

The circuit begins by placing one electron in each single-particle state beneath the Fermi level using X gates. Following this, XX+YY and P gates enact an orbital rotation into a basis that diagonalizes a subset of the Coulomb interactions $J_k$. The phases associated with $exp(iJ_k)$ that cannot be addressed within the orbital rotation operator would then be enacted by all-to-all connected CP gates in the UCJ algorithm. As the circuit in Fig.~\ref{fig:figS1} shows the local version LUCJ, these CP gates are pruned to match the honeycomb connectivity of recent IBM quantum devices. For simplicity, we have omitted two CX gates that would be necessary to enact the CP gate between the spin sectors (coupling q0 and q4) which are buffered by an ancillary qubit on real IBM hardware. These CP operations are followed by a further orbital rotation to project onto the original single-particle basis, on which determinants are sampled for exact diagonalization via the Davidson method.

Parameters are assigned from CCSD t2-amplitudes using the ffsim function $ffsim.UCJOpSpinBalanced.from\_t\_amplitudes()$, and the LUCJ operator circuit itself is generated using $ffsim.qiskit.UCJOpSpinBalancedJW()$ as in the Ref.~\cite{UCJ_tutorial} tutorial.

For the quantum simulation performed on the Quantinuum H2-2 quantum computer, we consider a system with size $L = 6$, which is mapped onto $24$ qubits, and use circuits with expansion orders $r = 1$ and $r = 2$. While only r=1 data are shown in the main text, the r=2 data show qualitatively identical energy trends and are included in Fig. S2(b) for comparison. The r=2 circuit has a significantly higher gate count, reflected in greater error (compare Fig. S2(c-d)). Since the Quantinuum H2-2 machine has 56 qubits, we stack (i.e., take the tensor product of) two identical 24-qubit circuits to form a 48-qubit circuit, effectively doubling the number of shots per circuit execution on the quantum hardware. The data are collected from a total of $3500$ hardware shots for each of the $r=1$ and $r=2$ circuits, resulting in effectively $7000$ shots for each 24-qubit circuit.

\begin{figure}[h]
    \centering \includegraphics[width=0.6\textwidth]{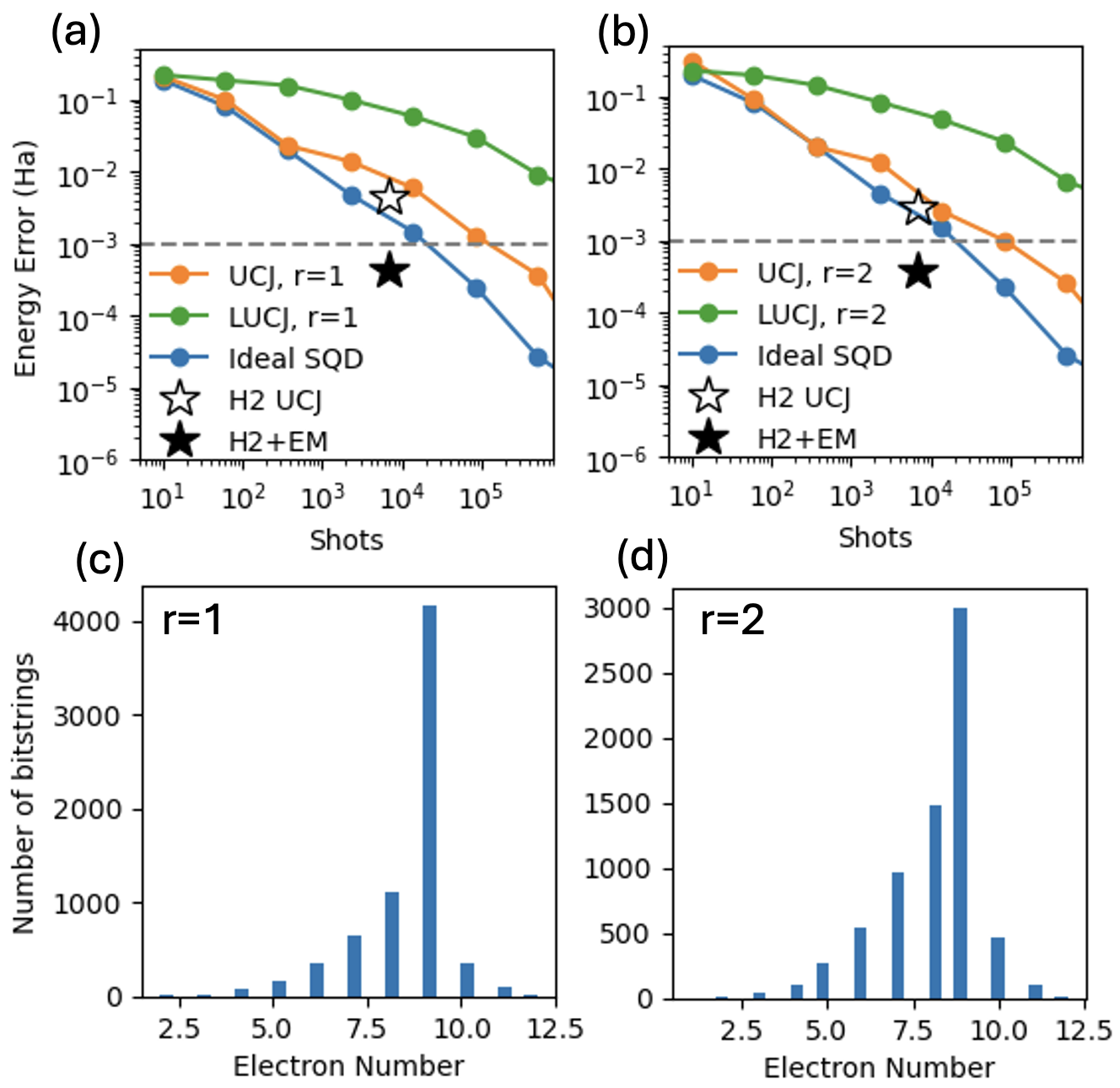}
    \caption{\textbf{Quantinuum data.} (a-b) Energy vs shot number curves of (a) r=1 and (b) r=2 calculations for the L=6 chain, as in Fig. 3(d) of the main text. Energies obtained from uncorrected Quantinuum data are plotted with hollow stars, and error mitigated energies are plotted with solid stars. (c-d) A histogram of spin up electron number in non-error-mitigated determinants from the Quantinuum H2-2 calculation. Error free calculations should have 9 electrons, and greater error in the r=2 calculation reflects an approximately doubled circuit depth.}
    \label{fig:figS2}
\end{figure}

\section{Defining the chain Hamiltonian}

The two-band model cuprate chain Hamiltonians are obtained via a multi-step procedure with two purposes:

1. Firstly, to preserve the first-principles-derived property that 2-body Coulomb interactions must be represented in the form
\begin{equation}
\mathbf{H}_{\alpha\beta\gamma\delta} = V_{\alpha\beta\gamma\delta}c^\dagger_\alpha c^\dagger_\beta c_\gamma c_\delta 
\end{equation}
and cannot be fully diagonalized on any real space basis as

\begin{equation}
\mathbf{H}_{\alpha\beta} = V_{\alpha\beta} c^\dagger_\beta c_\beta c^\dagger_\alpha c_\alpha, 
\end{equation}
where $\mathbf{H}$ is the interacting Hamiltonian, V is a constant, and $c^\dagger_\alpha$ ($c_\alpha$) is an electron creation (annihilation) operator acting on some spin-orbital basis. The scenario in Eq. (2) is a simplification adopted in the popular Hubbard model, and would allow the Hamiltonian to be fully represented in the r=1 order of UCJ.

2. Secondly, to provide a plausible model for the cuprate spin chain compound SrCuO$_2$. To accommodate the minimal two-band orbital basis in our model, correction terms are applied to the kinetic Hamiltonian and long-range Coulomb interactions to more closely approximate the experimentally identified spin excitation energies and charge density distribution.
\\ \\
\textbf{First principles dimer calculations}

The Hamiltonian for the dimer molecule in Fig. 1(a, top) of the main text is prepared via first principles Hartree-Fock (HF) calculations using the pyscf library with a 6-31g* basis for a charge neutral plaquette with formula Ca$_5$Cu$_2$O$_7$, using a Cu-O nearest neighbor distance of 2.1 $\AA$ and a Ca-O distance of 2.0 $\AA$. These atomic coordinates are chosen to resemble SrCuO$_2$, which is a model cuprate spin chain compound \cite{Shen_SCO112_occupancy, CKim_spinons_2006}. A minimal four-state active space is then chosen within the HF basis by selecting the low energy sigma-bonding and anti-bonding molecular orbitals of Cu $3d_{x^2-y^2}$ electrons on the CuO plaquette as in the two-band model of cuprates \cite{twoBand1, twoBand2, twoBand3, ZhangRice1988, twoBand_Sawatzky}.

This active space and the associated 1- and 2-body Hamiltonian terms are then projected onto a tight binding $|p\rangle \oplus |d\rangle$ orbital basis, with one orbital per atom. This is done to simplify orbital-resolved analysis and the extension of these Hamiltonians to a chain of variable length. The procedure used is as follows:
\\ \\
\textbf{Two-particle Hamiltonian terms}

1. Effective local orbital symmetries $|2p_\sigma\rangle_{eff}$ and $|3d_{x^2-y^2}\rangle_{eff}$ are read out from the HF states, as superpositions of $\sigma$-oriented O $2p$ and $3p$, and Cu $3d_{x^2-y^2}$ and $4d_{x^2-y^2}$ atomic orbitals, respectively.

2. A mapping from the HF active space $|\Psi_\alpha\rangle_{HF}$ to the tight binding orbitals is defined as:

\begin{equation}
|\Psi_\alpha\rangle_{TB} =(\Sigma_{i} |p_i\rangle\langle 2p_{\sigma,i}|_{eff} + \Sigma_{j}|d_j\rangle\langle 3d_{x^2-y^2,j}|_{eff}) |\Psi_\alpha\rangle_{HF} / norm,
\label{eq:tb_mapping}
\end{equation}
where $norm$ is the L$_2$ normalization factor, $|p_i\rangle$ and $|d_i\rangle$ are tight binding orbitals, and $i$ and $j$ index the planar oxygen and copper atoms, respectively.

3. The 2-body Coulomb Hamiltonian $H_C = \frac{1}{2}\sum_{\alpha,\beta,\gamma,\delta} C_{\alpha,\beta,\gamma,\delta} c^{\dagger}_{\alpha} c^{\dagger}_{\beta} c_{\delta} c_{\gamma}$ is mapped from the HF basis to the chain basis by interpreting the creation and annihilation operators $c^{\dagger}_{\alpha}$ and $c_{\alpha}$ as referencing the $|\Psi_\alpha \rangle_{TB}$ state basis. Terms that index sites in both plaquettes are further multiplied by a factor of 0.5 to coarsely represent screening effects that are not accounted for due to the minimal active space.
\\ \\
\textbf{Single-particle Hamiltonian terms}

A minimal 3-parameter model is defined for the kinetic Hamiltonian, consisting of on-site $|p\rangle$ and $|d\rangle$ orbital energies and a term for nearest neighbor Cu-O hopping. These parameters are obtained via the optimization of a fidelity-based loss function:

\begin{equation}
L = \sum_{i,j} w_{i,j} \left(1 - |\langle\psi_j|\phi_i\rangle|^2\right).
\end{equation}

Here, $|\phi_i\rangle$ and $|\psi_j\rangle$ are eigenstates of the first principles kinetic Hamiltonian and the fitted tight binding model, respectively. The inner product is calculated by first representing the states $|\phi_i\rangle$ on the tight binding basis as in Eq.~(\ref{eq:tb_mapping}). However, we note that the eigenstates of the kinetic Hamiltonian are not equivalent to the HF single particle states referenced in Eq.~(\ref{eq:tb_mapping}), which factor in a mean-field treatment of Coulomb Hamiltonian.

The weights $w_{i,j}$ are proportional to a pwhm = 1 eV Lorentzian function of the energy difference $L = (\frac{1}{(E_i-E_j)^2 + (0.5 eV)^2})$, and are normalized to sum to one along the tight binding wavefunction axis ($\Sigma_j w_{i,j} = 1$). The resulting kinetic Hamiltonian is further downscaled tuned by a multiplicative factor of 0.7 to match the spin chain energetics of SrCuO$_2$ (see discussion below). The need for tuning of this sort is unsurprising given the minimal HF active space.
\\ \\
\textbf{Extension to arbitrary-length chains}

The tight binding model is used to generate the single-particle kinetic Hamiltonian of chains with arbitrary length. The dimer-derived two body interaction terms described above are likewise used to define two electron interactions within the same plaquette and between neighboring plaquettes. For L$>$2 chains, the inequivalence of the central and terminal oxygens of the chain creates an ambiguity in how these interactions should be mapped. To maximize homogeneity, we therefore average all interaction terms that can be achieved through the superposition of a dimer onto any section of the chain. Lastly, we minimize the expectation value of the Hamiltonian on a single-Slater-determinant basis to identify the Hartree-Fock single-particle basis states of the new Hamiltonian.
\\ \\
\textbf{Characterization and fine tuning}

As noted above, inter-plaquette Coulomb interactions are reduced by 50$\%$ to coarsely account for the effect of screening, which involves higher energy orbital degrees of freedom not present in our model. Furthermore, the kinetic Hamiltonian is downscaled (multiplied) by a factor of 0.7 to roughly align spin excitation energies with expectations based on the experimentally observed Heisenberg spin interaction of $J=-0.23$ eV for SrCuO$_2$

An ideal Heisenberg spin chain with length L=2, 4, 6 and $J=-0.23$ eV has excitation energies of E = 0.230, 0.152, and 0.113 eV respectively, which have similar amplitude to the values of E $=$ 0.099, 0.180, and 0.138 eV for our corresponding cuprate chains. The low value observed for L=2 chains appears to be due to the fact that each plaquette of this chain sees Coulomb interactions from just a single neighbor. Setting interplaquette Coulomb interactions to 0 yields excitation energy ratios expected for an ideal spin chain (e.g. E(L=2)/E(L=6)) to within 2$\%$ but is not compatible with our goals for the paper. Defining the Hamiltonian in this way yields a charge density distribution that matches expectations for SCO, as described with respect to Fig.~\ref{fig:fig1}(c) of the main text.

\section{Sampling and averaging methods}

Determinant vs shot number curves in the main text display the expectation value of determinant number rather than the mean output of a fixed number of measurements. This expectation value is obtained by assuming the determinant distribution from a single $10^{6.5}$ shot measurement to represent accurate probabilities for the occupancy of each individual determinant.

A similar approach is adopted for the simulation energy curves: determinants for up to $10^{5.5}$ shot measurements are sampled from the output bitstring distribution of a single $10^{6.5}$ shot measurement. Plotted energies represent the average over 10 sampling batches. 

The Quantinuum H2 data in Fig.~\ref{fig:figS2} and Fig.~\ref{fig:fig3}(d) of the main text are not processed in this way, as they were obtained from separate 500-shot runs on the quantum device as described in Section I above.

\section{Kinetic and HF+ single particle bases}

To create the HF+ basis, we begin with the single-particle eigenstates of the kinetic Hamiltonian $|\alpha\rangle_{kin}$. We then define a mixing matrix M such that the eigenstates of M on the $|\alpha\rangle_{kin}$ basis are the Hartree-Fock states. The HF+ basis is obtained as the eigenstates of a matrix M' with off-diagonal matrix elements doubled as

\begin{equation}
M'= 2M-diag(diag(M)).
\end{equation}

This choice was motivated by the observation that our Hartree-Fock and HF+ states maintain a significant correspondence with the kinetic basis states, such that the HF+ basis can be thought of as realizing an amplified version of the mean field corrections from Coulomb interactions. The off-diagonal matrix elements of M are in all cases smaller than the difference between the corresponding on-diagonal matrix elements ($M_{\alpha,\beta\neq\alpha}<|M_{\alpha,\alpha} - M_{\beta,\beta}|$), as can be seen in the example below of the Hartree-Fock mixing matrix M for the L=4 chain: \\

\noindent{
\[
\begin{bmatrix}
-1.00 & 0 & -0.008 & 0 & 0 & -0.049 & 0 & 0.426 \\
0 & -0.665 & 0 & -0.046 & -0.065 & 0.001 & 0.472 & 0 \\
-0.008 & 0 & -0.445 & 0 & 0 & 0.004 & 0 & -0.029 \\
0 & -0.046 & 0 & -0.141 & 0.019 & -0.001 & -0.097 & 0 \\
0 & -0.065 & 0 & 0.019 & 0.157 & -0.001 & -0.096 & 0 \\
-0.049 & 0.001 & 0.004 & -0.001 & -0.001 & 0.449 & -0.001 & -0.062 \\
0 & 0.472 & 0 & -0.097 & -0.096 & -0.001 & 0.66 & 0.001 \\
0.426 & 0 & -0.029 & 0 & 0 & -0.062 & 0.001 & 1.00
\end{bmatrix}
\]
}

\section{Energy convergence for all r- and L-values}

\begin{figure}[h]
    \centering \includegraphics[width=0.8\textwidth]{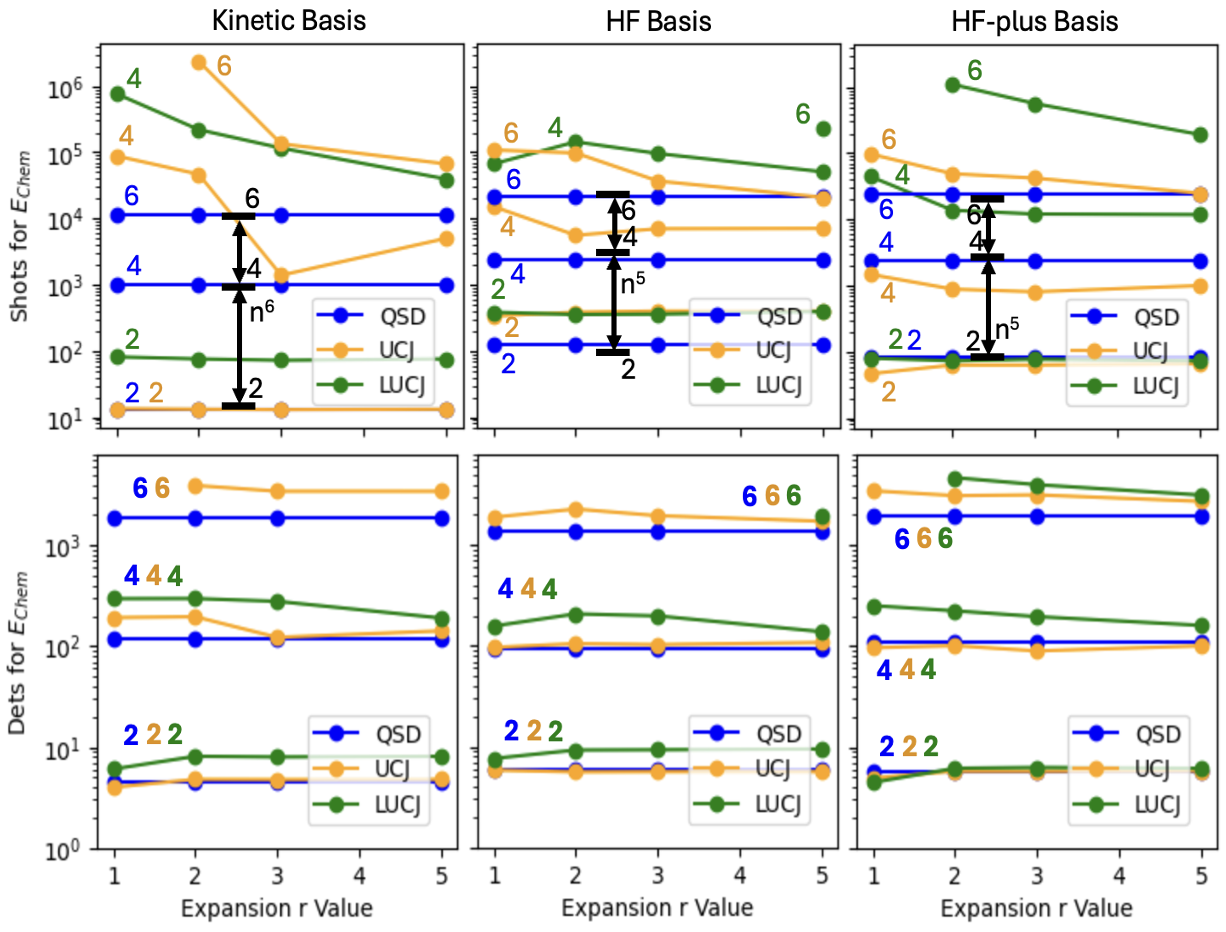}
    \caption{\textbf{Convergence to chemical accuracy in all calculations.} (top) Shots required to reach chemical accuracy for all molecular orbital bases and r-values. The chain length (L=2, 4, or 6) is indicated with color-matched numbers. Black brackets show vertical spacings expected for the indicated time complexity, and are aligned with data points for ideal SQD (labeled "SQD"). (bottom) Number of unique determinants in each basis at chemical accuracy.}
    \label{fig:figS3}
\end{figure}

The point of energy convergence to chemical accuracy is shown in Fig.~\ref{fig:figS3} for all combinations of single particle basis, UCJ expansion order (r) and chain length (L) considered in the paper. These systematics expand on Fig.~\ref{fig:fig4}(b) of the main text.

\end{document}